\documentstyle[aps,twocolumn,epsfig,amssymb]{revtex}
\def\ave#1{\langle #1\rangle}

\newcommand{\ve}[1]{{\vec #1}}

\begin{document}
\title{Anomalous Heat Conduction in a Di-atomic One-Dimensional Ideal Gas} 
\author{Giulio Casati$^1$ and Toma\v z Prosen$^2$}
\address{
$^1$International Center for the Study of Dynamical Systems, Universita'degli 
Studi dell'Insubria, I-22100 Como, Italy and Istituto Nazionale di Fisica della 
Materia and INFN, Unita' di Milano, Italy\\
$^2$Physics Department, Faculty of Mathematics and Physics, 
University of Ljubljana, Jadranska 19, SI-1000 Ljubljana, Slovenia}

\date{\today}
\draft
\maketitle
\begin{abstract}

We provide firm convincing evidence that the energy transport in a one-
dimensional gas of elastically colliding free particles of unequal masses is 
anomalous, i.e. the Fourier Law does not hold. Our conclusions are based on the 
analysis of the dependence of the heat current on the number of particles, of 
the internal temperature profile and on the Green-Kubo formalism.
\end{abstract}


Understanding the dynamical origin for the validity of 
the Fourier law of heat conduction in deterministic one-dimensional
particle chains is one of the oldest and most frustrating problems in 
non-equilibrium statistical physics\cite{history,casati85,PR}.
Due to some very basic unresolved issues the problem has been a source of
many recent publications\cite{recent,hatano99,PC00,Dhar01a,Garrido,LLPrep}.

In the absence of analytical results, these papers are  mainly oriented towards 
a numerical analysis of the problem. However, due to the delicate nature of the 
questions under discussion, numerical results sometimes lead to different 
conclusions. This is the case, for example, of the 1d hard-point particles with 
alternating masses for which opposite conclusions have been reached 
\cite{hatano99,Dhar01a,Garrido}. This disagreement is not extremely surprising since 
this system lies in the foggy region which separates clear, regular integrable 
systems, from the totally chaotic, deterministic, motion. Indeed this system has 
zero Lyapounov exponent and therefore it lacks of the exponential local 
instability which characterizes chaotic systems. On the other hand it has been 
shown \cite{Prozent} that such systems can exhibit Gaussian diffusive behaviour 
and, more recently \cite{Casatig}, an example has been shown of a system with 
zero Lyapounov exponent which however obeys the Fourier law. From the point of 
view of a general theoretical understanding, the fact that the alternating mass 
problem lies in this critical region renders particularly important to 
establish, beyond any reasonable doubt, its conducting properties. This is what 
we set up to do in the present paper.

We consider a one-dimensional gas of interacting particles with coordinates 
$q_n$ and momenta $p_n$ for which the hamiltonian can be written in the form
\begin{equation}
H = \sum_{n=0}^{N-1} h_n,\quad h_n = \frac{p_n^2}{2m_n} + V(q_{n+1}-q_n).
\label{eq:ham}
\end{equation}
The energy current from site $n$ to site $n+1$ is defined as
$ j_n = \{ h_{n+1},h_n\} $ and satisfies
the continuity equation $(d/dt)h_n = \{ H,h_n\} = j_n - j_{n-1}$.
In particular we focus our attention on the ideal gas model of elastically
colliding particles, $V(q>0) = 0, V(q<0)=\infty$, with alternating masses, 
$m_{2n-1}=m_1=\sqrt{r}$, $m_{2n}=m_2=1/\sqrt{r}$, 
where the ratio $r=m_1/m_2$ serves as
a model parameter. We have mainly considered the value $r=(\sqrt{5}-1)/2$;
however all the reported numerical results have been checked also
for several other values of $r$ ($0 < r < 1$) where we found no qualitative 
distinction.

We place our system of $N$ particles
between two stochastic maxwellian heat reservoirs at temperatures $T_{\rm L}$
and $T_{\rm R}$ (see \cite{casati85} for a description of the reservoir
model). 
We chose the temperatures of the reservoirs, $T_{\rm L}=1$
and $T_{\rm R}=2$, and measure the long-time averaged heat current
$\ave{J} = \lim_{t\to\infty}(1/t)\int_0^t dt' J(t')$ versus the
system size $N$, where  $J=(1/N)\sum_{n=1}^{N-1} j_n$.
Here we want to note that strict equivalence between our definition of the heat current 
(which simply accounts for the energy transferred during collisions), and the
`free particle' current $j_n = m_n v_n^3/2$ used by some 
authors, e.g \cite{Garrido} (which does not 'feel' the collisions),
is nontrivial in general \cite{LLPrep}.

In order to ensure that our results are not affected by finite size effects we 
have put particular care in using an efficient numerical scheme which 
allows to reach high N values. Our algorithm, developed for the first time in 
Ref.\cite{PR}, searches in a partially ordered tree ({\em heap}) of pre-computed 
candidates pairs for the next collision and, due to this, its requires only $\log_2 N$ 
computer operations per collision. As a consequence, we were 
able to simulate very long chains and we have obtained reliably 
converged results for lattices with sizes $N$ up to 30000.
Convergence has been controlled by checking the constancy of
the finite-time-averaged heat current $(1/t)\int_0^t dt' J(t')$, and to this 
end simulations for the largest system sizes had to be carried on up to $10^{12}$ 
pair collisions.
It is also clear that convergence problems suggest to keep far away form
the $r$ values too close to one or to zero. The analysis made in 
\cite{casati85} indicated that the range $0.15 < r < 0.6$ was the most effective in
attenuating solitary pulses and the value $r=0.2$ was chosen. In the
present paper we take the somehow 'standard' choice $r=(\sqrt{5}-1)/2$.

\begin{figure}
\vspace{-2mm}
\psfig{figure=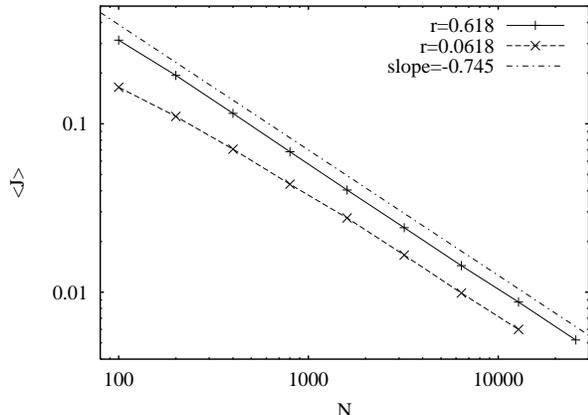,width=2.3in,angle=-90}
\caption{
Time-averaged energy current of a system of $N$ particles between heat
baths at temperatures $T_{\rm L}=1$ and $T_{\rm R}=2$ vs. the size $N$,
at two different mass ratios $r=m_1/m_2=
(\sqrt{5}-1)/2$ and $r=(\sqrt{5}-1)/20$.
The suggested scaling $\ave{J} \propto N^{-\alpha}$ with $\alpha=0.745$
is shown for comparison.
}
\label{fig:1}
\end{figure}

Validity of Fourier law implies the scaling 
$ J \propto \nabla T \propto N^{-1}$. Our numerical results shown in 
fig.~\ref{fig:1} clearly demonstrate instead a different power-law behaviour  
namely $J \propto N^{-\alpha}$ with $\alpha \approx 0.745\pm 0.005$ over a
very large range in $N$.
We have also found that the scaling exponent $\alpha$ does not change appreciably 
with the mass ratio $r$. For example for ten times smaller value of $r$ the 
asymptotic scaling only sets in later (i.e. for larger values of $N$, see 
fig.~\ref{fig:1}). The possibility of a slow convergence to the asymptotic value 
might be at the origin of the slightly different numerical values for $\alpha$ 
found in previous numerical experiments ($\alpha \approx 0.65$ in Hatano 
\cite{hatano99}, $\alpha \approx 0.83$ in Dhar \cite{Dhar01a}). 
Since the model under consideration is energy scaling we do not expect any dependence of the 
exponent $\alpha$ on the reservoirs temperatures.

The above results therefore solve the existing controversy and clearly show that 
the alternating mass, 1d hard point gas does not obey Fourier heat law. 
We turn now to the analysis of other quantities which, besides providing 
additional confirmation of the above conclusions, illuminate interesting
aspects of the heat conduction problem. 
A quantity of main interest is the internal local temperature
profile $T_n = \ave{p_n^2/(2 m_n)}$ in the non-equilibrium steady state for
the system placed in between the heat reservoirs. First we notice that the 
temperature profile in discrete 
index variable $n$ is different than the temperature profile in position 
variable $q_n$ \cite{Dhar01a} since the inverse density 
$dq/dn=\ave{q_{n+1}-q_n}$ is non-uniform in non-equilibrium, in fact it is
simply proportional to the temperature due to constancy of pressure
\cite{Dhar01a}.
Now, in case of a Fourier law, the thermal conductivity $\kappa$ scales with 
temperature like $\kappa \propto \sqrt{T}$. Therefore, from 
$\sqrt{T} (dT/dn)dn/dq = {\rm const}$ one obtains the temperature profile 
$T^{\rm kin}_n = (T_{\rm L}^{1/2}+(T_{\rm R}^{1/2}-T_{\rm L}^{1/2})n/N)^2$.
Extensive numerical simulations showed that the temperature profile in our model 
converges, for sufficiently large $N$, to a well-defined scaling function $T^{\rm scal}_n = f(n/N)$  
 which is slightly, but {\em significantly 
different}, from the kinetic temperature profile $T^{\rm kin}_n$.
This is another evidence for the anomalous heat transport and for the 
non-validity of the Fourier law in our system. It should be remarked that the 
convergence to the temperature profile predicted by kinetic 
theory observed in \cite{Dhar01a}, which has indeed been considered as surprising by the 
author himself, actually does not take place. 

A standard theoretical analysis of transport laws is based on Kubo formulae
\cite{kubo,Visscher74}.
However, applicability of  Kubo formula in momentum conserving cases,
i.e.  for translationally invariant systems like model (\ref{eq:ham}), is not very clear. 
This is particularly critical in view of a recent claim \cite{PC00} that Kubo formula 
diverges for a momentum conserving lattice with non-vanishing pressure. 
For this latter type of models we have an additional difficulty in applying the Kubo
formalism, namely, as we show below, the result depends not only on the
temperature gradient, but also on other thermodynamic properties
of the initial non-equilibrium state -- i.e. the
{\em isobaric} case
(constant pressure profile) or the {\em isodense} case 
(constant density profile). There is no {\em a-priori}
argument which favours either of these two options: the choice depends on the 
specific physical situation of interest. For instance the steady-state heat 
current simulation considered above (figs. 1,2) is clearly described by the 
isobaric state. Since we want to consider both situations we need to
revise the derivation of Kubo formula by 
following the time evolution of a general non-equilibrium initial state in an 
isolated system with periodic boundary conditions 
$q_N \equiv q_0 + N,p_N\equiv p_0$.
To this end, we prepare the initial state in a local-equilibrium
state described by the inverse temperature profile $\beta_n$ and by an additional
thermodynamic potential $\gamma_n$
\begin{equation}
\rho_{\rm neq} = Z^{-1}_{\rm neq} 
\exp\left(-\sum_n \beta_n h_n - \sum_n \gamma_n 
(q_{n+1}-q_n)\right)
\label{eq:Zneq}
\end{equation}

\begin{figure}
\vspace{-2mm}
\psfig{figure=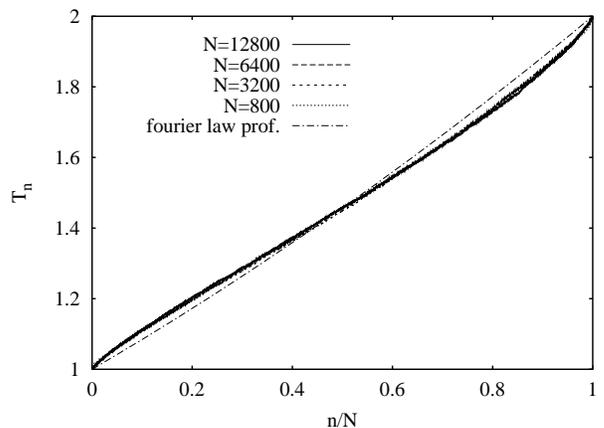,width=2.3in,angle=-90}
\caption{
Temperature profile for $T_{\rm L}=1$, $T_{\rm R}=2$, and different
sizes $N$ from $800$ to $12800$ (dotted-dashed-solid curves), as compared to the Fourier law prediction (chain
curve).
}
\label{fig:t}
\end{figure}

\noindent
This (small) additional term  is necessary in order to 
equilibrate the pressure in the isobaric case. 
Notice that $\gamma_n$ is undetermined up to an arbitrary additive
constant due to a gauge invariance of the second term in (\ref{eq:Zneq}).
In order to determine the gradient of the $\gamma$-potential
which is necessary to keep the physical pressure $\phi$ constant ($n$-independent), 
we compute the {\em generalized pressure} $\phi_l$
\begin{eqnarray}
\beta_l \phi_l &=& -\frac{\partial}{\partial a}\ln Z_l(a)\vert_{a=0}, \\
Z_l(a) &=& 
\int e^{-\sum_n(\beta_n V(q_{n+1}-q_n + a\delta_{ln}) -
\gamma_n (q_{n+1}-q_n + a\delta_{ln}))} d\ve{q} \nonumber
\end{eqnarray}
By a simple trick, a shift of one variable $q_l \to q_l + a$ in
the integral $Z_l(a)$, we find $Z_l(a) \equiv Z_{l-1}(a)$ and therefore
\begin{equation}
\beta_l \phi_l = \beta_{l-1} \phi_{l-1} = {\rm const}.
\end{equation}
Writing the {\em physical pressure} (force) as
$\phi = -\ave{V'(q_{n+1}-q_n)}_{\rm neq} = \phi_n + \gamma_n/\beta_n$,
multiplying by $\beta_n$, and taking the first difference 
$\nabla f_n := f_n-f_{n-1}$ we obtain the required 'gradient' 
\begin{equation}
\nabla \gamma_n = \phi \nabla \beta_n.
\label{eq:isobaric}
\end{equation}
 
In the following we consider two specific cases: (i) 
The initial isodense state with $\ave{q_{n+1}-q_n}_{\rm neq}={\rm const}$. This 
is obtained by putting $\gamma_n\equiv 0$.
(ii) The initial isobaric state with uniform pressure profile. This is obtained 
by specifying the $\gamma$-potential according to eq. (\ref{eq:isobaric}).
We note again that the isobaric state (ii) is the one which is
relevant for the steady non-equilibrium state of a system in contact with 
heat reservoirs. Carefully repeating the first few steps in the 
derivation of the Green-Kubo formula (following Ref.\cite{Visscher74}) we arrive at 
the very general linear response formula
$$
\ave{A(t)-A(t_0)}_{\rm neq} = \int_{t_0}^t dt' 
\ave{A(t')\sum_n(\nabla\beta_n j_n + \nabla\gamma_n v_n)}_{\rm eq}
$$
where $v_n = \dot{q}_n$ are the particles' velocities.
In the last step we have assumed that we are close to equilibrium
($\nabla\beta_n$ and $\nabla\gamma_n$ small) so that the RHS can be evaluated
in the corresponding equilibrium state $\ave{}_{\rm eq}$.
Let us now consider the periodic temperature profile
$\beta_n = \beta + \epsilon\sin(2\pi k n/N)$,
and compute the total heat that has been transported in time 
$t$ from {\em warm} to {\em cold} regions of the lattice, 
namely $Q(t) = \int_0^t dt' J_k(t)$ where $J_k := N^{-1}\sum_{n=0}^{N-1} \cos(2\pi k n/N) j_n$.
Inserting 
$Q$ for $A$ and using (\ref{eq:isobaric}) (case (i) is
obtained by formally setting $\phi=0$) we obtain 
\begin{equation}
\ave{Q(t)}_{\rm neq} = \frac{2\pi\epsilon}{N}\int_0^t dt'(t-t')
\ave{J_k(t)(J_k + \phi V_k)}_{\rm eq}.
\label{eq:Kubo}
\end{equation}
where $V_k := N^{-1}\sum_{n=0}^{N-1} \cos(2\pi k n/N) v_n$.
We see that the growth of the transported heat $\ave{Q(t)}_{\rm neq}$ is
given by the generalized correlation function
$C_k(t) = C_{JJ}(t) + \phi C_{JV}(t)$, where $C_{JJ}(t)=\ave{J_k(t)J_k}_{\rm eq}$ and
$C_{JV}(t)=\ave{J_k(t)V_k}_{\rm eq}$. In the isodense case (i) expression (\ref{eq:Kubo})
reduces to the usual current autocorrelation function only.

\begin{figure}
\vspace{-2mm}
\psfig{figure=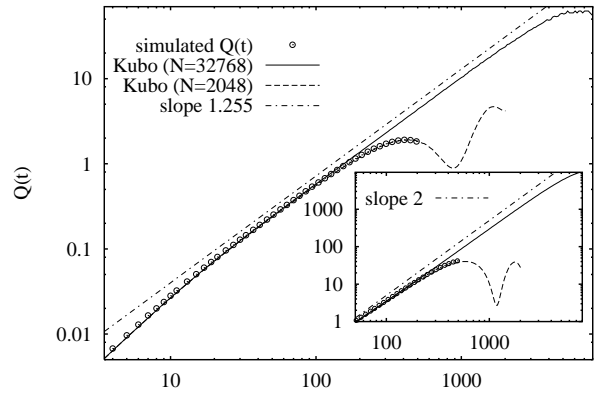,width=2.3in,angle=-90}
\caption{
Transported heat $Q(t)$ in an isolated system of size $N=2048$ 
obtained by starting from a non-equilibrium {\em isobaric} initial state (circles)
with $\epsilon=0.2$. For comparison we show the corresponding equilibrium
averaged Kubo-like expressions (\ref{eq:Kubo}) for $N=2048$ (dashed) and
for $N=32768$ (solid curve, multiplied by 16 to account for scaling 
$\ave{Q(t)} \propto 1/N$). The dashed-dotted line has a slope $1.255$ and gives
the best fit in the range $20 < t < 2000$.
The corresponding data for the {\em isodense} case are shown in the inset
with the ballistic slope $2$.
}
\label{fig:4}
\end{figure}

In the case of Fourier law, we expect initial linear growth
$\ave{Q(t)}_{\rm neq} \approx \epsilon \kappa t$,
while for the {\em ballistic} heat transport we expect
quadratic growth $\ave{Q(t)}_{\rm neq} \propto t^2$
(this has been confirmed numerically 
for the integrable gas of equal masses $r=1$).
However, in a generic system with momentum conservation and 
non-vanishing pressure $\phi\neq 0$, like our diatomic hard-point gas,
we find qualitatively different behavior in cases (i) and (ii).
For example, due to the result \cite{PC00}, $C_{JJ}(t)$ has a plateau $\propto \phi^2$ and the transport is 
ballistic in the isodense case, while in the isobaric case the second term,
$C_{JV}(t)$, compensates for the plateau and yields a much slower increase of the 
transported heat. In this latter case, independent numerical computations of
$\ave{Q(t)}_{\rm neq}$ and of $C_k(t)$ for
$N$ up to $32768$ shown in fig.~\ref{fig:4} give 
$\ave{Q(t)}_{\rm neq} \propto t^\nu$ with $\nu\approx1.255$, 
which is still clearly super-diffusive, and directly validate the formula 
(\ref{eq:Kubo}).

\begin{figure}
\vspace{-2mm}
\psfig{figure=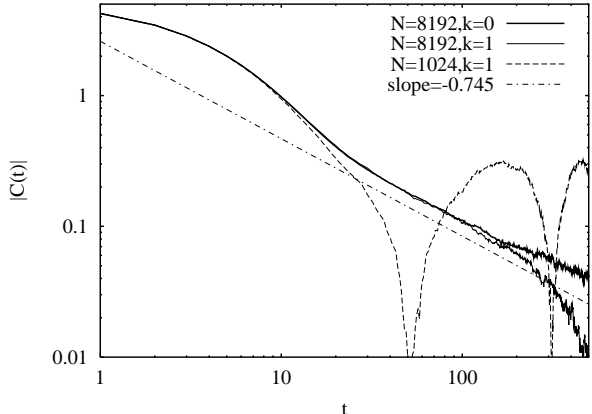,width=2.3in,angle=-90}
\caption{
The generalized time correlation function (see text) computed with canonical
average at $\beta=1$ for two system sizes $N=1024$ and $N=8192$ and for the 
zeroth $k=0$ (dashed/solid curve) and the first $k=1$ (thin curve) spatial 
Fourier mode. 
Note the $t^{-0.745}$ decay
(dashed line) in the range $20 < t < 200$ (for $N=8192$) 
whereas for longer times we see finite size effects (e.g. we have periodic 
oscillations for $k=1$ 
due to transversals of sound waves).
}
\label{fig:2}
\end{figure}

In fig.~\ref{fig:2} we show the generalized correlation function 
$C_k(t)$  for $k=0$ and $k=1$ separately. 
Note that the results for $k=1$ exhibits some
oscillations for longer times due to finite size effects, namely
due to periodicity of the lattice, while $k=0$ gives the {\em spatially 
homogeneous} correlation function which has the same long-time 
behavior with weaker finite size effects [however, the case $k=0$ is not 
strictly related to Kubo formula (\ref{eq:Kubo})]. We see that in the time range
where $C_0(t)$ and $C_1(t)$ match, the asymptotic decay is compatible with 
$C_k(t) \propto t^{-\mu}$ with the exponent $\mu = 2 - \nu = 0.745$ 
consistent with eq. (\ref{eq:Kubo}), and satisfying $\mu=\alpha$. 

\begin{figure}
\psfig{figure=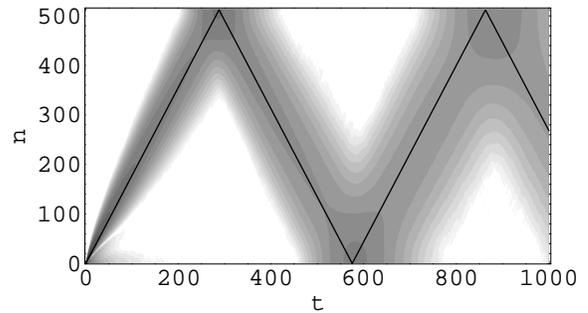,width=3in,angle=0}
\caption{
The spatio-temporal current-current correlation function $c_{jj}(n,t)$ at
temperature $1/\beta=1$ on a lattice of size $N=1024$ is shown with 
twenty different shades of greyness spaced equidistantly from $10^{-4}$ to 
$4.0$ in logarithmic scale. The zig-zag solid line indicates the peak 
ballistic sound-wave contribution moving with a uniform sound velocity $c_s=1.78$. 
}
\label{fig:3}
\end{figure}

This results can be interpreted in the following way. In the isodense
initial state the initial temperature gradients drive the heat ballistically 
in terms of sound waves \cite{PC00}. On the other hand, in the isobaric 
initial state, we have density gradients which drive the 
heat in the opposite direction and almost exactly compensate 
for the effect of temperature gradients so that the net effect is a 
sub-ballistic, but still super-diffusive, energy transport.
In order to illustrate the mechanism of ballistic energy transport we show in 
fig.~\ref{fig:3} the spatio-temporal current-current correlation 
function $c_{jj}(n,t) = \ave{j_0 j_n(t)}_{\rm eq}$ which exhibits clear 
ballistic tongues along the lines $n=\pm c_s t$ where $c_{\rm s}=1.78$.

In this paper we have discussed the thermal conducting properties of a one
dimensional hard point gas with alternating masses. This problem has a long
history and recently several numerical results have appeared which lead to 
contrasting conclusions on whether Fourier law is obeyed or not. 
In the latter case, different values have been found for
the rate of divergence of
coefficient of thermal conductivity as a function of the particles number.
Indeed, the slow convergence properties which sometimes characterize this problem
suggest particular care in the interpretation of numerical findings.

Here we have presented accurate numerical analysis, made possible by a
powerful integration scheme, which allows us to establish definite convincing
evidence that the system does not obey the Fourier law. Moreover, by considering 
a typical mass ratio $r=(\sqrt{5}-1)/2$,
we have found that the 
asymptotic scalings: (i) steady-state heat current between heat baths 
$\ave{J} \propto N^{-\alpha}$, (ii) heat transported within a non-equillibrium 
isobaric initial state in isolated system $\ave{Q(t)} \propto t^{2-\alpha}$, and
(iii) generalized current-velocity correlation in equilibrium state
$C_k(t) \propto t^{-\alpha}$, are described by just one exponent 
$\alpha=0.745$.

This work has been partially supported by EU Contract N0. HPRN-CT-1999-
00163(LOCNET network) and by MURST (Prin 2000, {\it Caos e localizzazione in 
Meccanica Classica e Quantistica.} TP acknowledges financial support by the
Ministry of Science, Education and Sport of Slovenia.

\begin{references}
\bibitem{history} M. Toda, Phys. Rep. {\bf 18}, 1 (1975);
W. M. Visscher, Meth. Comput. Phys. {\bf 15}, 371 (1976);
R. A. MacDonald and D. H. Tsai, Phys. Rep. {\bf 46}, 1 (1978);
E. A. Jackson, Rocky Mountain J. of Math. {\bf 8} 127 (1978);
G. Casati, J. Ford, F. Vivaldi, and W. M. Visscher,
Phys. Rev. Lett. {\bf 52}, 1861 (1984);
J. Ford,  Phys. Rep. {\bf 213}, 271 (1992).
\bibitem{casati85} G. Casati, Foundations of Physics {\bf 16}, 51 (1986).
\bibitem{PR} T. Prosen and M. Robnik, J. Phys A: Math. Gen. {\bf 25}, 3449 (1992).
\bibitem{recent} 
S. Lepri, R. Livi, and A. Politi, Phys. Rev. Lett. {\bf 78}, 1896 (1997); 
S. Lepri, R. Livi, and A. Politi, Physica D {\bf 119}, 140 (1998);
B. Hu, B.-W. Li, and H. Zhao, Phys. Rev. E {\bf 57} 2992 (1998);
D. Alonso, R. Artuso, G. Casati, and I. Guarneri, Phys. Rev. Lett. {\bf 82}, 1859 (1999);
C. Giardina, R. Livi, A. Politi and M. Vassalli, Phys. Rev. Lett. {\bf 84}, 
2144 (2000); O. V. Gendelman and A. V. Savin, Phys. Rev. Lett. {\bf 84},
2381 (2000); C. Mejia-Monasterio, H. Larralde and F. Leyvraz, 
Phys. Rev. Lett. {\bf 86}, 5417 (2001); A. Dhar, Phys. Rev. Lett. {\bf 86}, 
5882 (2001).
\bibitem{hatano99} T. Hatano,  {\it Phys. Rev. E} {\bf 59}, R1 (1999).
\bibitem{PC00} T. Prosen and D. K. Campbell, Phys. Rev. Lett. {\bf 84}, 2857 
(2000).
\bibitem{Dhar01a} A. Dhar, Phys. Rev. Lett. {\bf 86}, 3554 (2001).
\bibitem{Garrido} P.L. Garrido, P.I. Hurtado, and B. Nadrowski, 
{\it Phys. Rev. Lett.} {\bf 86}, 5486 (2001).
\bibitem{LLPrep} S. Lepri, R. Livi and A. Politi, {\tt cond-mat/0112193}.
\bibitem{Prozent} G. Casati and T. Prosen, Phys. Rev. Lett. {\bf 85}, 4261 
(2000); see also {\it ibid.} {\bf 83}, 4729 (1999). 
\bibitem{Casatig} Baowen Li, G. Casati, Lei Wang and Bambi Hu, in preparation.
\bibitem{kubo} R. Kubo, {\it J. Phys. Soc. Japan} {\bf 12}, 570 (1957).
\bibitem{Visscher74} W. M. Visscher, Phys. Rev. A {\bf 10}, 2461 (1974).
\end{references}
\end{document}